# Multi-exposure diffraction pattern fusion applied to enable wider-angle transmission Kikuchi diffraction with direct electron detectors

Tianbi Zhang (0000-0002-0035-9289), T. Ben Britton[*] (0000-0001-5343-9365)

Department of Materials Engineering, The University of British Columbia. 309-6350 Stores Road, Vancouver, BC, V6T 1Z4 Canada

* Corresponding author: ben.britton@ubc.ca

**Abstract**

Diffraction pattern analysis can be used to reveal the crystalline structure of materials, and this information is used to nano- and micro-structure of advanced engineering materials that enable modern life. For nano-structured materials typically diffraction pattern analysis is performed in the transmission electron microscope (TEM) and TEM diffraction patterns typically have a limited angular range (less than a few degrees) due to the long camera length, and this requires analysis of multiple patterns to probe a unit cell. As a different approach, wide angle Kikuchi patterns can be captured using an on-axis detector in the scanning electron microscope (SEM) with a shorter camera length. These 'transmission Kikuchi diffraction' (TKD) patterns present a direct projection of the unit cell and can be routinely analysed using EBSD-based methods and dynamical diffraction theory. In the present work, we enhance this analysis significantly and present a multi-exposure diffraction pattern fusion method that increases the dynamic range of the detected patterns captured with a Timepix3-based direct electron detector (DED). This method uses an easy-to-apply exposure fusion routine to collect data and extend the dynamic range, as well as normalise the intensity distribution within these very wide (>95°) angle patterns. The potential of this method is demonstrated with full diffraction sphere reprojection and highlight potential of the approach to rapidly probe the structure of nano-structured materials in the scanning electron microscope.

## 1. Graphical Abstract:

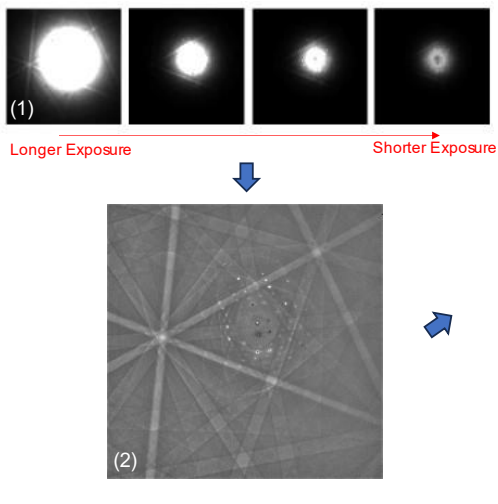
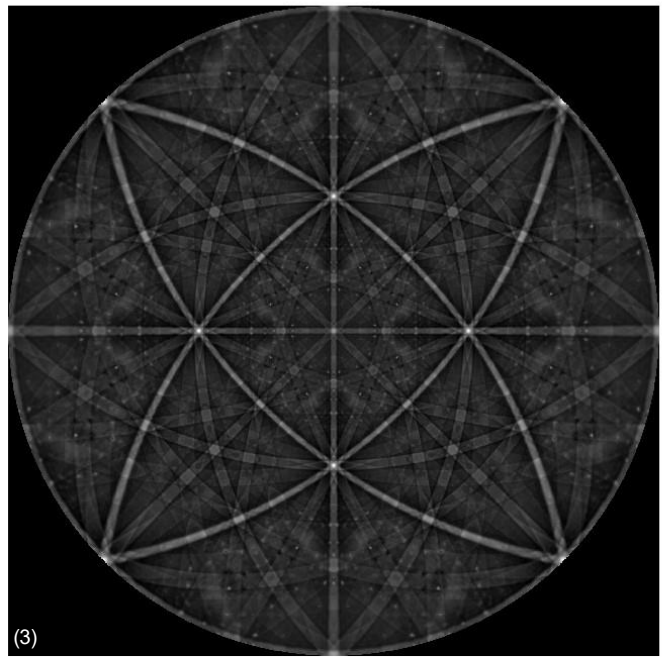

(1) Capture multiple Transmission Kikuchi Diffraction (TKD) patterns with a Direct Electron Detector
(2) Fuse & flatten multi-exposure TKD pattern
(3) Use crystal symmetry & alignment to generate full experimentally obtained stereographic sphere

## 2. Introduction

Transmission Kikuchi Diffraction (TKD) in scanning electron microscope (SEM) is an emerging orientation microscopy technique which was first proposed as 'transmission EBSD' in 2012 by Keller and Geiss [1]. TKD in SEM combines the high spatial resolution through scanning transmission electron microscopy in SEM (STEM-in-SEM) of transmission electron microscopy (TEM) thin foils, with the high angular resolution of Kikuchi diffraction-based techniques (such as electron backscatter diffraction, EBSD), ease of diffraction pattern analysis using conventional EBSD algorithms, together with the high accessibility of a SEM. Since it was first demonstrated, the TKD in SEM method has been widely used to enable new applications to characterize nanocrystalline materials and a comprehensive review of TKD in SEM applications can be found in the review by Sneddon et al. [2].

Initial TKD experiments mapped samples using a new sample holder to enable the conventional EBSD detector to be used in the so-called 'off-axis' geometry as the EBSD detector was not on the optical axis of the SEM. The off-axis method is accessible due to the limited changes in existing conventional EBSD hardware and software. However, analysis can be challenging as these off-axis TKD patterns (TKPs) tend to share the following characteristics: (a) the diffracted electrons are scattered by large angles, typically ~90° from the primary beam, resulting in poor electron yield, low SNR and long dwell times per point; (b) the pattern centre is above the top of the detector screen, so the scatter angle per pixel in the detector varies strongly, causing varying SNR and contrast across the TKP; (c) there is very large band divergence for bands towards the bottom of the screen, as the hyperbolic Kossel cones which form the edges of the Kikuchi bands widen more as the diffraction vector is further from the pattern centre position in the gnomonic projection [3].

These challenges motivated the development of the 'on-axis' approach originally proposed by Fundenberger et al. [4], where a scintillator is placed horizontally below the specimen and on the optical axis of the microscope. This was achieved using a limited modification of a conventional EBSD camera, where the conventional indirect scintillator was rotated 90° and coupled via a mirror. The on-axis geometry improves the SNR of the TKPs significantly, resulting in faster scans; and reduces artifacts due to gnomonic projection and simplifies the analysis of the patterns tremendously, resulting in improved indexing [3].

Additionally, the solid angle subtended in the on-axis TKD technique is also typically higher than TEM-based orientation microscopy techniques which yield spot diffraction patterns, such as (scan) electron nanodiffraction [5] and scanning precession electron diffraction (SPED) [6]. The typically higher voltage used in most TEM results in Kikuchi bands which are very close together and challenging to interpret and the narrow capture angle available within the TEM result in difficult to index Kikuchi patterns, as compared to the wide angle and low voltage TKD patterns which can be analysed using conventional EBSD analysis routines [8]. Furthermore, the TKD method also has a wide capture angle which means that multiple interplanar angles are present within the (direct) projection of the unit cell as presented within the TKD pattern and this provides access to alternative analysis strategies, such as those used in for advanced EBSD pattern analysis [7–9] which improve the angular resolution of the technique [10].

The TKD approaches discussed so far have used conventional indirect EBSD detectors, where the signal containing diffracted electrons is first converted to photons by a scintillator, and then transferred to a digital camera (typically a charge coupled device (CCD) or a complimentary oxide-metal-semiconductor (CMOS) device) via optical couplings [11]. This two-step conversion-transfer process introduces inefficiencies, optical blurring, diffraction pattern distortions, and reduces the SNR of the detector [12]. Furthermore, this scintillator and CCD/CMOS camera assembly can have limited dynamic range.

For on-axis TKP analysis, there are three major issues associated with the very strong variation in electron dose incident on the screen:

- The presence of the direct beam, and a range of scattering events which result in a large electron dose for electron scattering angles <5°. Additionally, for thin samples, there can be spot diffraction patterns up to <15°. [13]
- The probability of electrons scattering decreases as the angle increases.
- Due to the gnomonic projection, at low scattering angles, the pixels subtend a wider angle and so the total dose per detector pixel is larger for low angles as compared to high angles, and this is important when using an electron counting device.

To address these challenges, direct electron detectors (DED) such as hybrid pixel array detector (HPD) or monolithic active pixel sensor (MAPS) can be employed [14], which are based on radiation-hardened solid-state detectors. For the present work, of particular interest is HPD based on the Medipix or Timepix chips operating in the event counting mode and frame readout [15,16] due to the larger electron flux allowed per pixel per unit time and large pixel size. These features enable the detector to capture a wide camera angle while being placed relatively far from the sample in the absence of a post-sample lens in most SEMs.

For high dynamic range experiments, it is worth noting that DEDs are able to detect single electron events, and they typically have better modulation transfer function, detection quantum efficiency and better SNR than indirect detectors [17,18]. Hence, DEDs should enable capture of higher quality patterns (i.e. higher SNR and higher optical acuity/sharpness), which concurrently enables the patterns to be analyzed with more sophisticated routines and allows comparison with high quality dynamical diffraction based simulations.

There have been reports in the literature which have demonstrated the ability to capture Kikuchi patterns generated by EBSD and TKD in SEM with DEDs [14,19]. These initial works have shown significant promise. However, there lacks an in-depth study of the raw TKD data captured by DED, and one particular challenge related to on-axis TKD-in-SEM remains. Although on-axis TKD theoretically has benefits for wider angle analysis compared to typical TEM-based analysis methods, as the intrinsically lower electron yield at higher scattering angles means that features at higher angular range will not be significantly recovered, even when a high SNR detector such as a DED is used since the SNR of an image is ultimately determined by the number of quanta (electrons) used to create the image for each pixel.

One method to mitigate this issue is to simply use longer exposure times (i.e. capture more electrons), but in practice there is a challenge associated with exceptional demands on the dynamic range requirements of the detector due to saturation, such as the 1022 counts per pixel per frame when a Timepix3 is operated in frame readout and event counting mode [20].

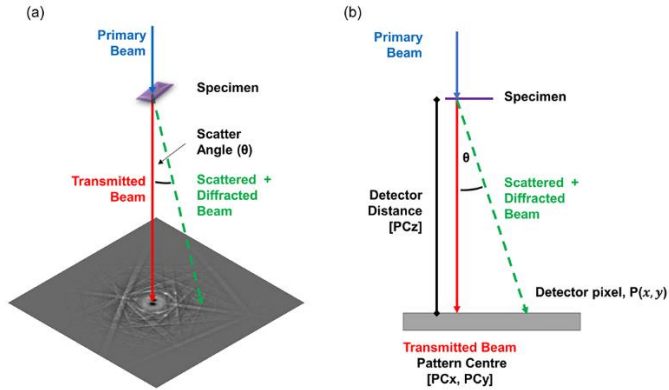

Figure 1: Schematic of the on-axis transmission Kikuchi diffraction geometry. (a) three-dimensional view with an example pattern; (b) side, two-dimensional view showing the relationship between the primary beam & pattern centre, specimen, and the scattered and diffracted beams in the pattern.

Therefore, although high angle features can be enhanced by increasing the exposure time, more pixels will likely be saturated around the direct beam, causing more loss of information as a trade-off. This issue is commonly seen in previously reported on-axis TKPs (e.g. in refs. [3,4]).

Some previous studies have attempted to mitigate the limited dynamic range of DEDs in STEM applications due to the direct beam. Mir et al. combined the two 12-bit readout counters of a Medipix3 detector, which effectively increases the dynamic range to $2^{24}$, and demonstrated a diffraction pattern with a dynamic range of approximately 1:3000 [21]. This approach can be useful but requires more sophisticated modifications to the read out structure (i.e. use both 12-bit channels to provide a 24-bit channel). Bammes et al. developed a hybrid counting strategy for STEM imaging with MAPS [22], which first identifies the sparse and non-sparse regions of the detector and then applies event counting and charge integration for the two regions respectively. However, due to the different architectures, sparse counting is often necessary for MAPS and this strategy cannot be readily adapted to HPD since HPD tends to accommodate a higher number of electrons per pixel before reaching saturation.

An inspiration for obtaining higher dynamic range TKPs in this work comes from a classic framework proposed by Debevec and Malik for obtaining high dynamic range photographs using light photography [23], which is further developed and termed 'exposure fusion' by Mertens et al. [24]. In this method, the scene is captured at a range of exposure times and the data is fused together with weighting based upon the exposure time. Variants of this method are commonly used in consumer cameras, including cell phones, to augment low light or high dynamic range scenes, e.g. so that a photographer can augment the image to compress the dynamic range prior to printing or representing in 8-bit greyscale or 24-bit color.

In this work, we present a data acquisition and processing routine for obtaining high dynamic range and therefore high angular range TKPs using a Timepix3-based HPD without modifying the detector architecture. The routine was developed through studying the variation of the signal within a pattern with respect to scattering angle (i.e., the illumination profile) and exposure time. In addition to the capturing of higher dynamic range patterns through multi-exposure diffraction pattern fusion, the fusion model enables us to calculate a simple model of the variation in background with scattering angle, and therefore provides a direct method of background correction based upon the physical signal capture. This normalises the intensity variation for all scattering angles, which significantly simplifies subsequent diffraction pattern analysis which is shown here when a single pattern is used to reconstruct the full spherical pattern.

## 3. Materials and Methods

**Sample Preparation.** Aluminum samples supplied by Rio Tinto Aluminium were taken from an extruded bar of an Al-0.71Mg-0.91Si-0.20Fe-0.003Mn wt.% alloy without heat treatment. Thin 3 mm disks were cut from the specimen and electropolished using a solution of 30% nitric acid (68% aqueous solution by weight) and 70% methanol by volume at -20 °C using a Struers Tenupol-5 twin-jet polisher to produce a sample with a hole and a nearby thin section which is suitably electron transparent.

**Microscope and Detector.** On-axis TKD was performed in a Zeiss Sigma field emission gun SEM at 30 keV electron energy and 300 pA beam current as measured using a Faraday cup and a built-in picoammeter.

A custom stage and a sample holder were designed and manufactured to enable routine access to our on-axis geometry (Figure 2). The sample holder clamps a TEM foil and is mounted on an XYZ hybrid linear piezoelectric stage (SmarAct GmbH, Germany) to position the sample above a MiniPIX Timepix3 HPD (Advacam s.r., Czech Republic) mounted below the stage. The stage was developed to be loaded in a conventional SEM chamber with minimal changes to the existing apparatus and is mounted onto the Zeiss Sigma base with a standard dovetail. The detector and stage are connected to the computer via vacuum feedthroughs.

In this work, a compact and (relatively) inexpensive MiniPIX platform based on Timepix3 was selected, as it has a large sensor area and features a simple USB 2.0-based readout [25] which is connected via a standard 4-pin LEMO vacuum feedthrough. The sensing layer is a 100 µm-thick Si sensing layer with no Al coating. As with all Timepix3-based detectors, the sensor is bump-bonded to the readout circuit with 256×256 pixels where each read out pixel is distributed within a square grid, which has a pitch of 55 µm. The sensing area is approximately 14×14 mm$^2$.

For the present work, we operate the detector in the in 'Event+iToT (integrated time-over-threshold)' mode with a 50 V bias, and patterns were obtained from event data. A minimum energy threshold of 3.01 keV was applied to supress readout noise according to factory calibration and verified by the authors. The detector contains 56 dead pixels (0.085% of total pixels), which are masked during capturing so they will register 0 event counts.

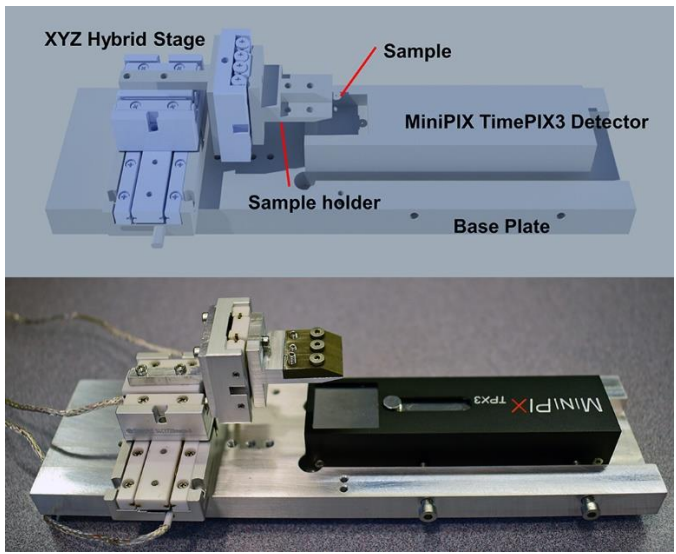

Figure 2: CAD rendering and photo of the custom TKD stage. Dimensions (W x H x D) of the base stage is 160×9×42 mm$^3$.

The stage was first aligned to place the direct beam towards the centre of the detector. In the on-axis configuration, this determines the X and Y positions of the pattern centre, and an initial measurement of the pattern centre, $[PCx, PCy]$, can be taken from observing the direct beam position with the detector. The detector distance (i.e. the Z component of the pattern center, $PCz$) was determined directly from the sample to detector distance. The position of pattern centre was verified and refined within the pattern matching step. Although only single patterns are presented here, the detector was also aligned within the scanning beam direction of the SEM by directly striking the primary beam on the detector while it is scanning in a raster pattern at low magnification (when the sample was moved out of the way) and observing the beam path via the detector read out.

Several grains were identified using the secondary electron detector in the SEM, and then the microscope was set to spot mode and controlled using an interface via the Zeiss SmartSEM API. This was connected (in software) to a simple python interface to control the detector and enable multi-exposure diffraction patterns to be captured while the beam was positioned on the same spot. A final pattern will be formed using the exposure fusion routine outlined in the next section. In this work, frames were collected at the following exposure times: [0.1,0.01,0.005,0.001] seconds per frame, and this makes the exposure total time per frame 0.116s. For this demonstration work, the presented data is the average of 50 frames. Collecting and averaging (or integrating) multiple frames result in higher pattern quality and SNR at a cost of increased collection time, as well as a more accurate estimation of the illumination profile of the patterns, which is a crucial step of the routine proposed in this work. Examples of single frame patterns are shown in the Appendix.

2.3 Processing and analysis of Kikuchi patterns

The goal of pattern processing here is to provide interpretable patterns that can be analysed to reveal information about the diffracting unit cell.

The processing routines proposed in this work are programmed in MATLAB within the AstroEBSD toolbox (https://github.com/benjaminbritton/AstroEBSD/) which has been updated to include these new routines.

Details of the new approach for on-axis multi-exposure pattern summing and flat fielding will follow, and subsequently flat fielded patterns were analyzed using the existing pattern matching routines using both ESPRIT DynamicS version 1.0.5977.45721 (Bruker Nano GmbH) and AstroEBSD using template matching and the refined template matching approaches [8,26]. For the pattern matching steps, dynamical Kikuchi patterns were simulated via DynamicS, which generates a reference 'cube' pattern using the method described by Winkelmann et al. [27,28]. The cube contains the intensity of the dynamically simulated EBSD pattern for each vector originating from a reference unit cell for a specified electron energy. Once the reference cube is generated, a *.bin file is saved to disk and then AstroEBSD can reproject this according to the geometry of the detector (i.e. pattern centre and pixel array size) and crystal orientation using a bicubic interpolation scheme in MATLAB. The refined template matching routine with pattern centre refinement provides sub-pixel accuracy of the pattern centre.

## 4. Theory of Image Processing for On-axis TKPs

Here we apply the multi-exposure fusion technique used in digital light photography, together with a flat field operation that removes the variation in electron scattering yield as a function of scattering angle. The proposed process of obtaining the high dynamic range TKP includes four steps: (1) probing raw event data as a function of scatter angle; (2) weighting of the raw event data; (3) fusing the weighted event data; and (4) flat fielding the fused pattern.

3.1 Raw Event Data as a Function of Scatter Angle

The flat detector is placed below the sample and the projection of the pattern can be described with regards to a known pattern centre $[PCx, PCy, PCz]$. The gnomonic projection of the pattern which originated from the sample can be used to calculate of the radial scattering profile, as a function of scattering angle θ:

$$tan\theta = \sqrt{(PC_x - x)^2 + (PC_y - y)^2} \times \frac{1}{PC_z} \times \frac{1}{256} \quad (1)$$

where each pixel position is described by the coordinates *(x, y)* in the 256-by-256 pixel array (as shown in Figure 1). Figure 3 shows example patterns with different exposure times and event profiles as a function of scatter angle.

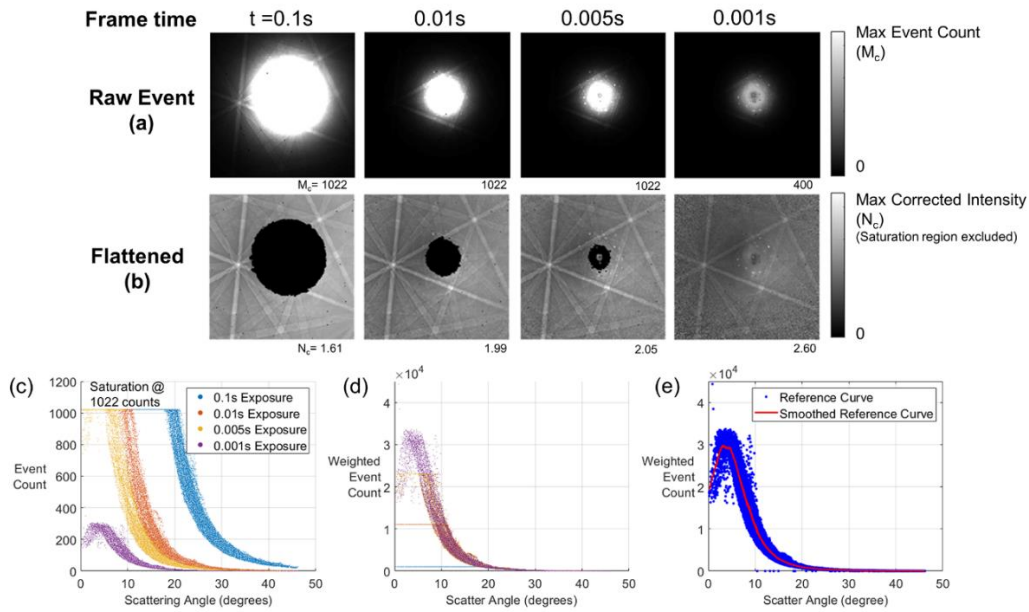

Figure 3: TKD patterns of an aluminum sample obtained at 30 keV electron energy and 0.001 s, 0.005 s, 0.01 s and, 0.1 s exposure times with 50 frames averaged. (a) Raw patterns, (b) corresponding flattened patterns (saturation regions assigned with 0 intensity). Pattern centre is [0.5454, 0.4350, 0.7579], corresponding to a maximum scattering angle of 50°. (c) Plot of event count of the patterns as a function of scattering angle of the raw aluminum dataset. (d) Plot of weighted signal as a function of scattering angle with a base time of 0.1s. (e) Plot of the radial smoothing function used for the flat field step. (Color figure online)

In Figure 3 (a), long exposure times result in saturation of the pattern near the direct beam but high-quality features at wide scattering angles. Shorter exposure times result in recovery of scattering information and features closer to the direct beam, but for these exposures there is limited scattering per frame at wider scattering angles.

Using these patterns and the knowledge of the pattern centre, radial integration of the 2D patterns was performed to create plots of event count as a function of scattering angle, and examples of these radial profiles from an example pattern are shown in Figure 3(c).

These profiles show the following trends:

1) At low scattering angles and long exposures, the pattern is saturated (at 1022 counts for this detector and read out).
2) For short exposure times, the scattering near the spot can be observed and there is a region of decreased intensity and a peak at 4°.
3) After this peak, the scattering profile smoothly decreases towards zero.
4) For short exposure times, the scattering profile at large angles is close to zero and dominated by noise.

Assessment of these profiles can be enhanced if we consider that they contain four potential signals:

1) A smooth 'background' of scattered electrons which varies predictably with the scattering angle from the direct beam. This may depend on sample thickness, probe shape, and incoming probe voltage.
2) The Kikuchi pattern, which is a fraction of the scattered electron signal and formed from diffracted electrons that escaped the foil.
3) Noise, which includes a wide range of contributions that are dependent on the: sample, sensor, detector readout, stray X-ray and electrons incident on the detector, as well as shot noise.

4) For samples which are suitably thin, like the example in Figure 3, there can be additional spot diffraction patterns around the direct beam. Note that due to high electron flux, some diffraction spots may appear dark due to coincidence loss, i.e., multiple electron events in quick succession that are counted as a single event.

Each radial scattering profile is extracted for the average of 50 exposures for each recorded exposure time. These profiles show that that event count increases at lower scattering angles until reaching the saturation limit of the detector (1022 events per frame) which occurs near to the direct beam. At larger scattering angles, the signal decreases towards zero.

The noise is important to separate into the two factors that result in a low SNR at wide scatter angles. The incident electron beam has a low probability of being scattered to these wide angles, and additionally the angle subtended per pixel is smaller in the gnomonic projection for wider scattering angles. Together these mean that if the exposure time is very short, there will not be enough electrons to generate the signal above the noise level within each frame, and so simply counting many short frames is not an effective way of recovering wide angle information as summing of many short exposure images can result in reproduction of the average noise.

In practice, the challenges of saturation and noise present two challenges:

- At low scattering angles, there are high incident electron counts on the detector. This means that for longer exposures, a saturation plateau with a value of 1022

- counts per frame occurs at lower scatter angles (where scattering information is lost), but the wide-angle signal increases in intensity and SNR improves.
- At wide scattering angles there are low incident electron counts on the detector, and for short exposures the noise dominates (even when the signal is summed over a very large number of frames). However, these short exposures reveal that the direct beam (and any associated low angle scattered beam) may not saturate the detector. Further information is shown as the structure of the pattern near the direct beam is complicated – either due to local count saturation, or other scattering and diffraction-based effects that we have qualitatively observed to vary with sample thickness and will be subject of further study.

To address this problem and produce a diffraction pattern that can be analysed for a range of scattering angles, we can identify the saturated areas and discard this information. For the other areas, we can weight the raw data according to the exposure time and fuse the frames together.

3.2 Weighting of Raw Event Data

Weighting is achieved by assuming that the response of a DED in event-counting mode (which detects single electron events) is linear with respect to the number of electrons detected up to the saturation limit, i.e. there is negligible variation in beam current and detector gain. Under this assumption, the amount of incident electrons as well as detected electrons should scale linearly with exposure time.

In this work, we weight the raw event data based on exposure time per frame. Here we consider $t_0$ as the longest exposure time per frame used in the capture sequence and subsequent processing. Therefore, for another exposure time per frame $t_n$, the ideal weighting factor should be the ratio $w_n = t_0/t_n$. However, considering shot noise in individual experiments, the actual scale factor should fluctuate around the ideal value, which we denote as $w_n = t_0/t_n + \Delta_n$. Thus, following the framework by Debevec and Malik for light photography [23], the general illumination profile of an on-axis TKD pattern can be obtained by collecting TKD patterns from the same spot on the sample (i.e. capturing the same scene) at different exposure times, and weighting the data against a baseline exposure time ($t_0$) according to Equation (2):

weighted data at exposure time $t_n$ = raw data at $t_n * w_n$ (2)

For a series of exposure times $t_1...t_n$ ($n > 0$), its corresponding weighted data can be treated as the general illumination profile in the region which is the difference between its own saturation region and the saturation region at $t_{n-1}$. Within this zone, weighted data of exposure time $t_n$ has the highest SNR in the dataset and is thus used for the subsequent fusion step. To determine the actual scaling factors ($t_0/t_n + \Delta n$), the ratios of pixel values at an exposure time $t_n$ in the unsaturated region at $t_0$ is calculated at each pixel and the average value is taken as the scaling factor for the dataset taken at $t_n$. for the datasets show in this paper, $|\Delta n|$ is within 10% of to $t_0/t_n$ .

Figure 3(d) shows a plot of the weighted data versus scatter angle for the demonstration dataset, where all weight data, except for the saturated pixels, follow approximately the same relationship described by the weighted data of the shortest exposure time (0.001 s in this case).

3.3 Fusing the Weighted Event Data

To obtain high angular and high dynamic range patterns the weighted event data is fused according to the following algorithm:

1) the longest exposure time is chosen as the baseline $t_0$ and the corresponding weighted data is used as the base pattern;
2) the saturated region of the base pattern is filled by weighted data of a shorter exposure time $t_1$ until reaching its own saturation region;
3) this process is repeated until the weighted data of the shortest exposure time (which ideally does not saturate) of the dataset is filled in the region corresponding to the smallest scattering angles.

A step for dead pixel correction is followed by replacing the 0 intensity values of dead pixels, which were masked during data acquisition, by corresponding pixel values of the 3×3 median filter of the raw fused pattern.

As shown in Figure 3(e), the fused pattern has a much higher dynamic range ($4\times10^4$ vs 1022 per frame) than a single exposed pattern. Due to the exposure fusion step, the equivalent exposure time of the fused pattern is higher than every single exposure pattern and here this is 1.16x the longest single exposure time.

3.4 Flattening the Fused Pattern

The raw fused pattern can be flattened to remove the 'background signal' associated with the change in scattering probability with scattering angle, prior to performing further Kikuchi pattern analysis. A data-informed, pattern-specific method is developed here using the general illumination profile of each TKP. For each pattern and the associated dataset, the weight event data captured at the shortest exposure time is taken as a 'reference curve', which describes the general relationship between scattering angle and the signal for this particular diffraction pattern, if the detector had a much higher dynamic range. The reference curve itself though is of low SNR, especially at high scattering angles.

The reference curve is smoothed by a Savitzky-Golay filter with a $2^{nd}$ order polynomial to obtain a radially averaged illumination profile of the TKP, and the raw fused pattern can be divided by this function, in effect providing a flat-field operation. An additional Gaussian high pass filter is applied to the demonstration pattern to correct for uneven contrast in the fused pattern. This Gaussian filter step is optional to make analysis even easier, but analysis can be performed without the filter and a comparison is shown in Figure A3 in the Appendix.

Note that once the general function is fitted, this smoothing function can be scaled according to each exposure time and used to flat-field single-exposed patterns as well as each individual frame in the dataset, if required (Figure 3(b)).

## 5. Results

An example dataset from the Al sample is used to demonstrate the effectiveness of the presented data processing routine.

The flattened fused pattern has a uniform contrast and shows Kikuchi bands with good contrast and sharpness at higher scattering angles up to ~50° (Figure 4), comparing to single exposed patterns especially at shorter exposure times, while effectively reducing the saturation domain and thus the information lost.

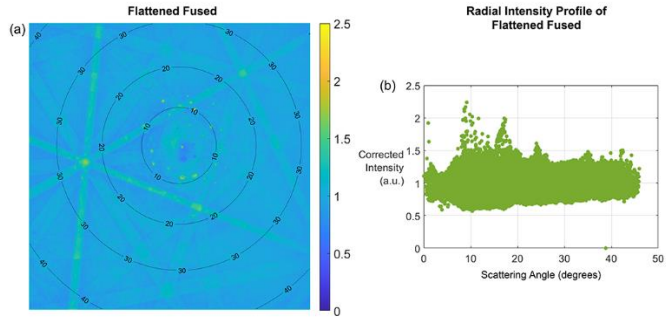

Figure 4: (a) The flattened fused pattern with overlaid contours of scatter angles; (b) radial profile of corrected intensity of the flattened fused patterns. (Color figure online)

Figure 5 shows the enhanced band contrast of the flattened fused pattern compared to other single-exposed patterns and the flattened dynamically simulated pattern for two selected bands. These two bands are in the zone formed by the longest exposure time in the dataset, and as a result of the processing routine their intensity profiles coincide. It is evident that the signal to noise ratio that provides band contrast improves with exposure time.

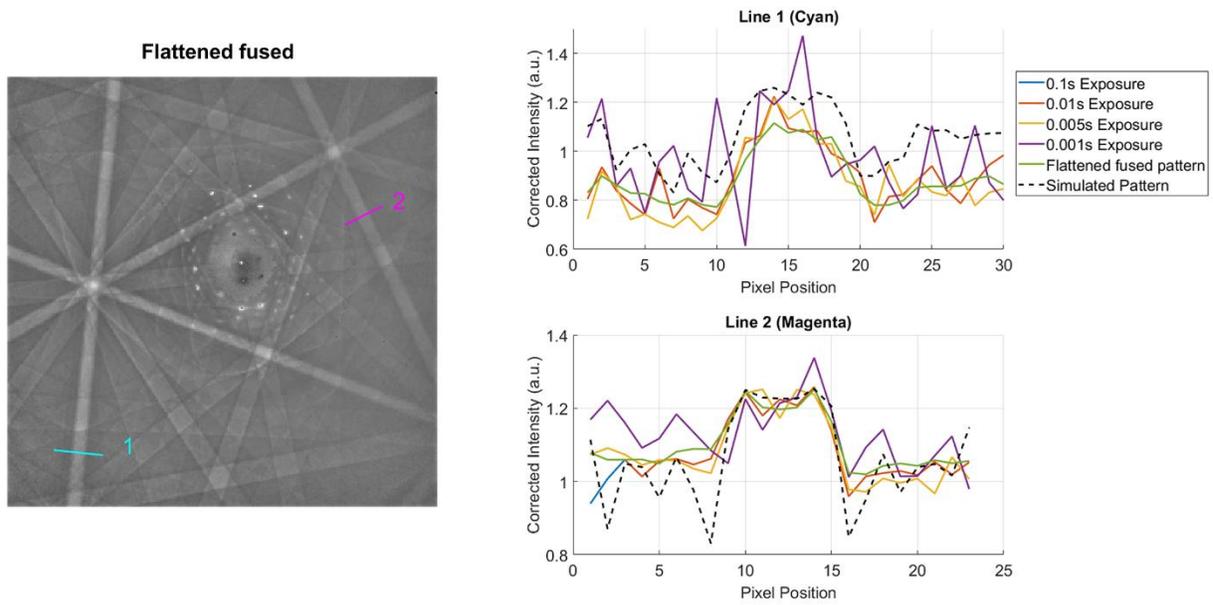

Figure 5: Line profiles across two Kikuchi bands (cyan and magenta) in the flattened fused pattern, raw single-exposed patterns, and the normalized dynamically simulated pattern of the example pattern. (Color figure online)

The enhanced angular range and contrast benefits both Hough transformation-based and pattern matching based indexing as reflected by the high-quality Hough peak map and the cross-correlation coefficient when matched with the dynamically simulated pattern (Figure 6).

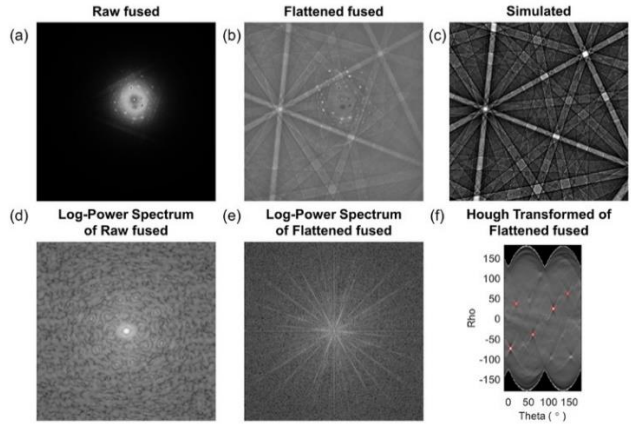

Figure 6: (a) Raw fused, (b) flattened fused pattern and (c) the matched dynamically simulated patterns of the example pattern (cross-correlation coefficient=0.728). (d) and (e) are log10 2D FFT spectra of (a) and (b) respectively. (f) Hough peak map of (b) and detected peaks. Indexed Euler angles (Bunge convention) are [142.06°, 23.15°, 244.98°] from pattern matching.

Figure 7 shows a superposition of the fused and flattened captured experimental pattern outlined in blue overlaid on the stereographic projection of a dynamical simulation (with the projection mathematics given in the Appendix). Due to the quality of the pattern, limited distortions and precise knowledge of the crystal orientation and camera geometry, there is extremely good alignment.

The individual pattern shown here contains multiple copies of the fundamental zone (one fundamental zone is shown in yellow in Figure 7(a)), and therefore can be used to 'inflate' the pattern, via crystal symmetry, to fill the full diffraction sphere (a detailed description of the method is outlined in refs [29–31]). In brief, the flattened fused pattern is re-projected onto the diffracting sphere and symmetry operators of the aluminum unit cell (24) are applied to rotate and overlap the patterns and the result is shown in Figure 7b. (The projection mathematics to convert from the sphere to the stereogram is detailed in the Appendix). The intensity of overlapping regions from different applications of the symmetry operators is normalized based on the number of symmetrically equivalent patterns overlapped for each point within the stereogram, shown in Figure 7c. The absence of blurring in this projection is due to the quality of the experimental pattern, and precise knowledge of both the crystal orientation and the detector geometry.

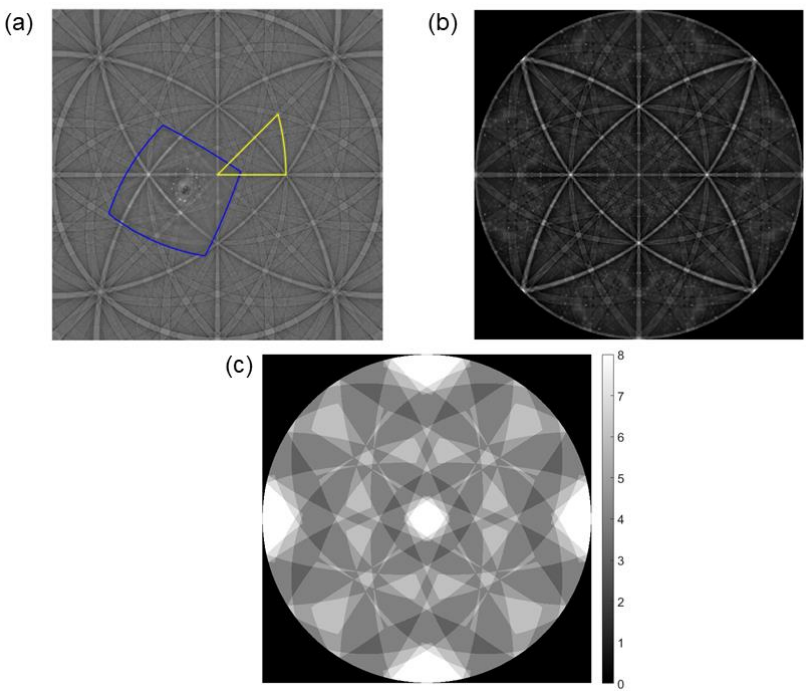

Figure 7: (a) Dynamical simulation of the stereogram of the diffracting sphere, with the flattened fused pattern (blue) reprojected and overlaid. The fundamental zone of the Al unit cell is shown in yellow. (b) Stereogram of the diffracting sphere reprojected using only the flattened fused pattern and the 24 cubic symmetry operators. (c) Map of the number of overlapped symmetrically equivalent patterns as a function of position on the sphere used to generate (b). (Color figure online)

A gallery of TKPs from aluminum processed by the routine in this work, along with corresponding simulated Kikuchi patterns is shown in Figure 8 to demonstrate the universality of the present method on patterns obtained from different crystallographic orientations. Note that these additional patterns do not contain spots and have different structure around the direct beam, which is probably related to variations in sample thickness for the electropolished region.

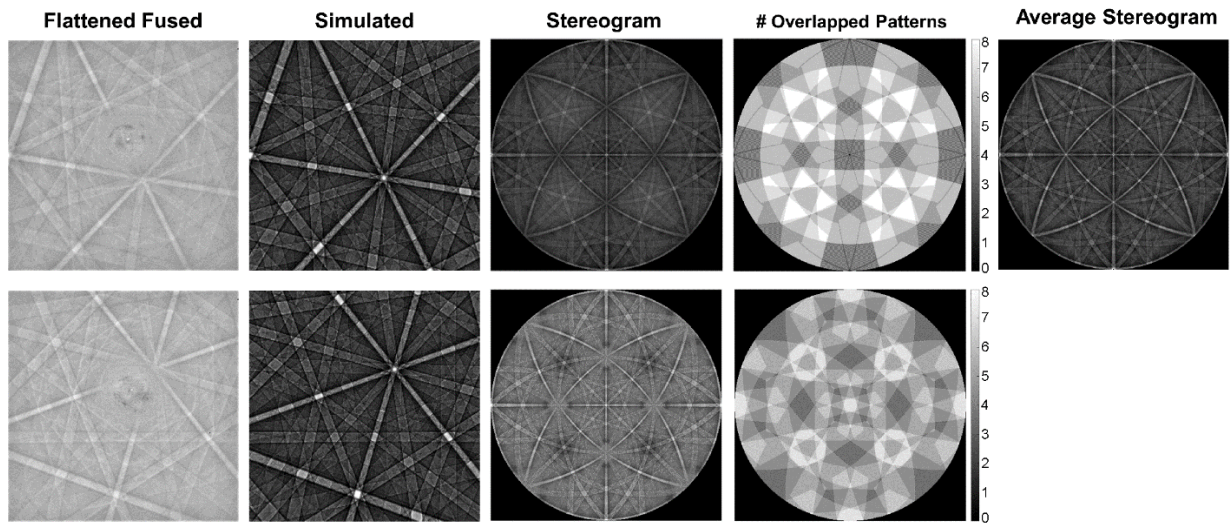

Figure 8: Additional experimental Al TKPs collected at 30 keV and processed using the routines described in this work. Left to right: flattened fused pattern, dynamically simulated pattern obtained from pattern matching, stereogram of the diffracting sphere using reprojected flattened fused pattern, map of the number of overlapped symmetrically equivalent patterns as a function of position on the sphere used to generate the stereogram, and the averaged stereogram of Figure 7(b) and together with those in 8.

## 6. Discussion

The major strength of TKD is that the Kikuchi patterns contain a wide angular range and look like conventional EBSD patterns, and therefore similar well-established methods can be used. Off-axis TKD has grown in popularity as there is very little modification to the SEM required, and conventional on-axis TKD requires minimal further modification while greatly improving the gain of the pattern, as more electrons are scattered and incident on a camera placed directly below the sample. In this work, we build on these strengths by presenting a DED-based system which fits in most SEMs with limited modifications, allowing features at very large scattering angles to be captured with lower beam currents. This is combined with a physics-based pattern normalization method for high pattern acuity and simple-to-use pattern analysis to enhance robust interpretation of on-axis TKD patterns significantly, through a combination of multi-exposure fusion to address challenges with significant variations in signal to noise ratio across the pattern. The flexibility in the detector and stage used in this study adds further degrees of freedom to the desired capture angle and the user can also opt to use a higher or lower probe current (combined with exposure time). In this work we opt to highlight the potential of capturing high angle patterns at a lower beam current.

Each TKD pattern from this approach has high signal to noise ratio for a significant angular range (±50° from the primary beam) which is unusual for a transmission electron microscopy method, and the scattering is centred on the direct beam. This means that each pattern describes many vectors in the unit cell, including both the in-plane and out of plane lattice vectors. We can imagine that there will be specific STEM-in-SEM applications where the user may wish to "zoom in" (by increasing the camera length) on particular features, e.g. to access fine structure information and increase the angular resolution of the pattern (perhaps at high scattering angles), or alternatively the user may wish to use a wide capture angle to reveal as much of the unit cell as they can in a single shot (as performed here). The pattern fusion method and pattern normalization approach are flexible and will benefit either analysis mode. Potential applications include strain or tetragonality measurements as well as phase classification (e.g. refs [32–34]).

Although this work only demonstrated an example dataset on aluminum, in theory this routine can be applied to any crystalline material where high spatial resolution is desired, provided there is enough diffraction volume for Kikuchi diffraction and the SNR of the raw patterns is acceptable. Other factors that may adversely affect pattern quality also include the use of supporting film for nanoparticles, which may cause additional beam broadening, pattern blurring [35] and irregularities in the radial event profile, and therefore, the thickness and evenness of the supporting film must be carefully chosen to reduce these effects. However, care must be taken in choosing an appropriate beam current for certain materials where beam damage could lead to amorphization. The high detection quantum efficiency of Medipix- or Timepix-based detectors can compensate the use of lower electron dose in terms of image quality [18].

Furthermore, this manuscript presents a new background correction method – via radial integration of the on-axis pattern, which is informed by the physics of scattering and enables flat fielding of the detected Kikuchi pattern especially at wide capture angles. A key step in obtaining the flattened fused pattern is to calculate the radial average of the illumination profile. Notably, there can be many ways to obtain an approximation of this profile by fitting or filtering the data. The main criterion used by the authors in selecting the method is that the value of the fitted or smoothed curve should be within the bounds of the original weighted data, in order to ensure an even background and contrast in the flattened fused pattern, and meanwhile maintain a reasonably fast execution of the MATLAB routine.

For thin enough samples, the resulting pattern often contains the spot and Kikuchi patterns. While we have observed the spot pattern here, we have not used the intensity variation of this spot pattern for any further analysis yet. We can imagine that combined knowledge of the precise electron dose, experiment geometry, scattering profile, spot pattern, and wide angle Kikuchi pattern presents the opportunity to significantly upgrade the quality of information we can obtain with S(T)EM-based experiments.

The exposure fusion routine is proposed as an approach to optimize the use of a lower dynamic range direct electron detector, such as the Timepix3 in frame readout mode, to achieve high signal to noise.  As the availability of small form factor Medipix3 detectors, and alternative detector architectures are made available, we imagine that there may be alternative and simpler strategies to achieve high dynamic ranges for these applications. This routine is specifically useful to address noise issues at high scattering angles for low probe currents, when there is limited signal to noise (see Figure 3b).

Furthermore, this manuscript presents a new background correction method – via radial integration of the on-axis pattern, which is informed by the physics of scattering and enables flat fielding of the detected Kikuchi pattern especially at wide capture angles to simplify interpretation of the diffraction signal with existing EBSD software analysis routines.

One challenge that this method does not directly overcome is the problem of coincidence loss [36]. In the Timepix3 architecture if the pixel is flooded with too many events in each counting time, then the measured number of incident events for that pixel can be significantly lower than the number of incident high energy particles. This problem is observed near to the direct beam in Figure 3(c). In the present work, this is addressed using masking and replacement with frames taken with shorter exposure times. In theory, an alternative detector could address this more directly and we are excited to see this potential if more small form factor detectors suitable for the compact design shown here are made available in the market.

## 7. Conclusions

In summary, this work presented a method to obtain high dynamic range on-axis TKPs with high angular ranges. Through studying the raw event data as a function of exposure time and scattering angle, a high dynamic range pattern can be fused together using time-weighted diffraction patterns captured at single exposure times. The result is a fused pattern with a dynamic range exceeding the single-frame dynamic range limitation of the Timepix3-based detector. The relationship between event data and scattering angle also enables an effective flat field routine to greatly enhance band contrast and angular range of the pattern, which is confirmed by robust indexing by Hough transformation, high quality pattern matching-based indexing, projection onto the stereographic projection of a dynamical pattern, and use of crystal symmetry to 'inflate' to describe the entire unit cell. This routine can be applied to a wide range of materials for high spatial and high angular resolution measurements.

**CRediT Authorship Contribution Statement**

**Tianbi Zhang**: conceptualization, investigation, methodology, data curation, formal analysis, software, visualization, validation, writing – original draft; **T. Ben Britton**: conceptualization, funding acquisition, project administration, resources, supervision, visualization, writing – review & editing.

**Acknowledgments**

We acknowledge the support of the Natural Sciences and Engineering Research Council of Canada (NSERC) [Discovery grant: RGPIN-2022-04762, 'Advances in Data Driven Quantitative Materials Characterisation']. We would like to thank Dr. Kirsty Paton (Paul Scherrer Institut), Dr. Ruth Birch (University of British Columbia) and Dr. Alex Foden (Imperial College London) for helpful discussions; Dr. Warren Poole, Dr. Gwenaëlle Meyruey (University of British Columbia) and Rio Tinto Aluminium for providing the aluminum alloy; and Mr. Berhard Nimmervoll and Mr. David Torok (University of British Columbia) for their help in designing and manufacturing the TKD stage. <mark>We thank Håkon Wiik Ånes for reading our preprint and informing us of an error with a figure that we have now corrected.</mark>

**Data Availability Statement**

Raw data is available on Zenodo (https://doi.org/10.5281/zenodo.8030030) and processing scripts are available as a part of the AstroEBSD software package (https://github.com/benjaminbritton/AstroEBSD/, and in the v1.1 release found here: https://doi.org/10.5281/zenodo.8078806).

**Appendix**

Figure A1 shows a comparison between single-frame patterns and patterns averaged over 50 frames, showing that single-frame patterns have a much higher level of noise.

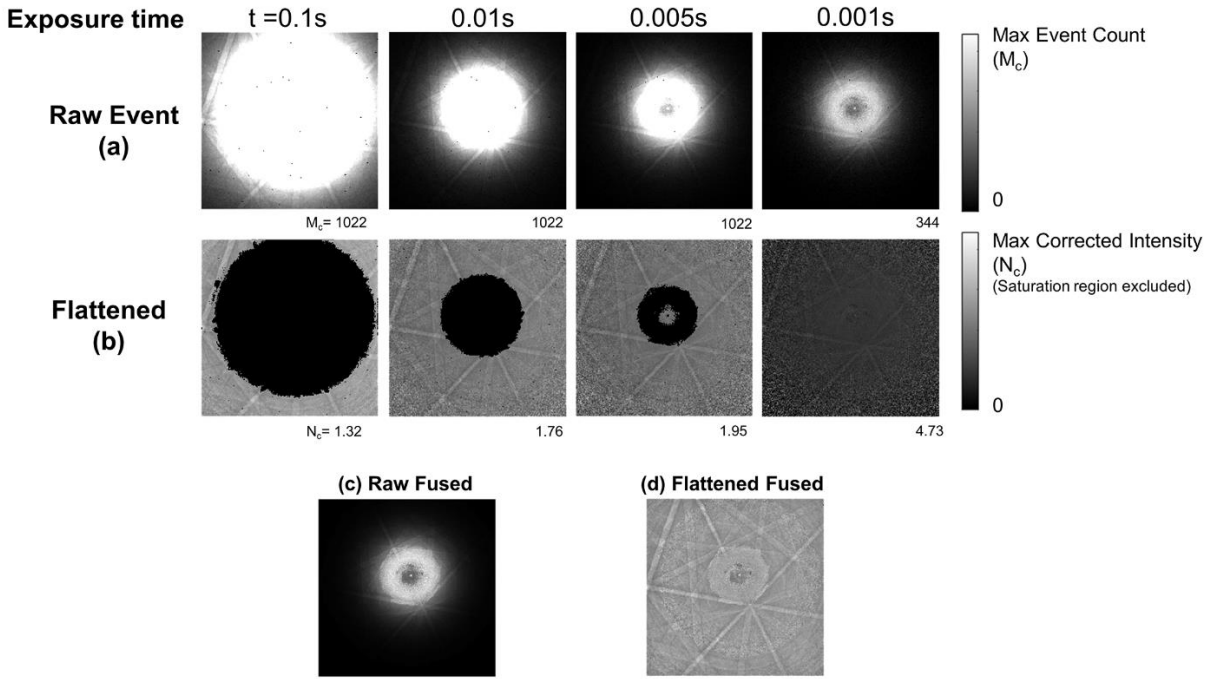

Figure A1: Single-frame TKPs of Al obtained at 30 keV. (a) Raw patterns, (b) background corrected patterns (saturated region assigned with corrected intensity of 0); (c) raw fused pattern and (d) flattened fused pattern.

Detailed steps for reprojecting the stereogram are outlined below. A schematic diagram of stereographic projection is shown in Figure A2.

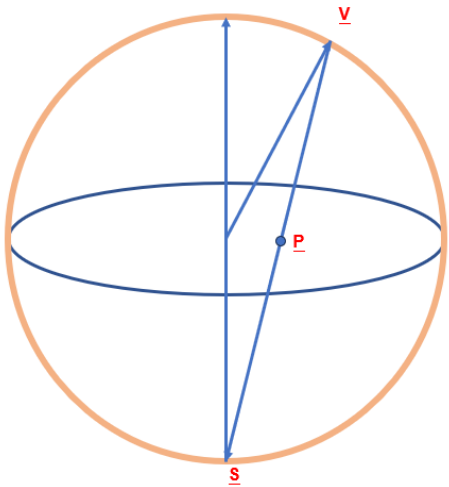

Figure A2: A schematic diagram of stereographic projection onto the equatorial plane to point $P$ for a vector, $V$, where $S$ is the south pole.

(1) From Figure A2, we can convert from a point on the sphere to a point on the equatorial stereogram via this vector equation:

$$P = S + \lambda(\overrightarrow{VS}) = \begin{bmatrix} 0 \\ 0 \\ -1 \end{bmatrix} + \lambda \begin{bmatrix} V_x \\ V_y \\ V_z + 1 \end{bmatrix}$$

We know that in the equatorial plane, $P_z = 0$, so can solve for lambda:

$$\lambda = \frac{1}{(1 + V_z)}$$

Therefore:

$$P = \frac{1}{(1 + V_z)} \begin{bmatrix} V_x \\ V_y \\ 0 \end{bmatrix}$$

(2) From Figure A2, we can convert from a point on the equatorial stereogram to a on the sphere via this vector equation:

$$V = S + v(\overrightarrow{PS}) = \begin{bmatrix} 0 \\ 0 \\ -1 \end{bmatrix} + v \begin{bmatrix} P_x \\ P_y \\ 1 \end{bmatrix}$$

Now we know that:

$$|V| = 1$$

So we can write:

$$V_x^2 + V_y^2 + V_z^2 = 1 = v^2(P_x^2 + P_y^2) + (v^2 - 2v + 1)$$

$$0 = v^2(P_x^2 + P_y^2 + 1) - 2v$$

$$v(-2 + v[P_x^2 + P_y^2 + 1]) = 0$$

So either $v = 0$ [this is the south pole solution, $S$], or:

$$v = \frac{2}{[P_x^2 + P_y^2 + 1]}$$

Therefore, $V$ can be written as:

$$V = \begin{bmatrix} 0 \\ 0 \\ -1 \end{bmatrix} + \frac{2}{[P_x^2 + P_y^2 + 1]} \begin{bmatrix} P_x \\ P_y \\ 1 \end{bmatrix}$$

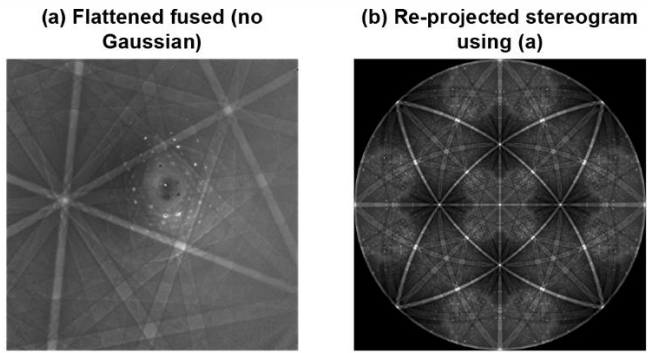

Figure A3: (a) The flattened fused demonstration TKP without the Gaussian filter correction. Its reprojection is shown in (b).